# The influence of magnetic vortices motion on the inverse ac Josephson effect in asymmetric arrays

Boris Chesca[a,1], Marat Gaifullin[a,b], Daniel John[a],

Jonathan Cox[a,c], Sergey Savel'ev[a], Christopher Mellor[d]

[a]Department of Physics, Loughborough University

Loughborough, LE11 3TU, United Kingdom

[b]SuperOx Japan LLC, Sagamihara, 252-0243, Japan

[c]School of Computer Science, University of Lincoln, Lincoln, LN6 7TS, United Kingdom

[d]School of Physics and Astronomy, University of Nottingham

Nottingham, NG7 2RD, United Kingdom

**Abstract**

We report on the influence a preferential magnetic vortices motion has on the magnitude of the inverse *ac* Josephson effect (the appearance of *dc* current Shapiro steps) and the coherent operation of asymmetrical parallel arrays of $YBa_2Cu_3O_{7-\delta}$ Josephson junctions (*JJ*) irradiated with microwave (*MW*) radiation in the presence of an applied magnetic field B. The preferential direction of motion of the Josephson vortices is due to the asymmetry-induced ratchet effect and has a dramatic impact: for a particular positive *dc* bias current *I* when the flux-flow is robust multiple pronounced Shapiro-steps are observed consistent with a coherent operation of the array. This suggests an efficient emission/detection of *MW* in related applications. In contrast, when we reverse the direction of *I*, the flux-flow is reduced and the Shapiro-steps are strongly supressed due to a highly incoherent operation that suggests an inefficient emission/detection of *MW*. Remarkably, by changing B slightly, the situation is reversed: Shapiro steps are now suppressed for a positive *I*, while well pronounced for a reverse current *-I*. Our results suggest that a preferential vortex-flow has a very significant impact on the coherent *MW* operation of superconducting devices consisting of either multiple *JJ*s or an asymmetrically biased single long *JJ*. This is particular relevant in the case of flux-flow oscillators for sub-terahertz integrated-receivers, flux-driven Josephson (travelling-wave) parametric amplifiers, or on-chip superconducting *MW* generators which usually operate at bias currents in the Shapiro step region.

[1] Corresponding author; email: B.Chesca@lboro.ac.uk

Achieving coherent microwave (*MW*) operation of Josephson junctions (*JJ*)-based devices operating in the presence of an applied magnetic field *B* is essential in many applications such as *MW* generators/detectors[1-2], Josephson flux-flow-oscillators (*FFO*) currently used as sub-terahertz integrated-receivers in radio-astronomical research or atmospheric science projects[3,4] or flux-driven Josephson (travelling-wave) parametric amplifiers[5-11] used in *MW* generation/detection[6-8] or read-out of a flux-qubit[9]. In all these applications, such *JJ*-based devices are *dc* current biased and usually operate in an environment where they are simultaneously exposed to both a (remnant) magnetic field *B* field and microwave (*MW*) radiation. The exposure to *MW* can be due to an external source or due to internally generated radiation via the *ac* Josephson effect. Depending on applications the presence of either *B* or *MW* can be essential for, or detrimental to, their operation. Such *JJ*-based devices are inherently asymmetrical to some degree (the asymmetry can be structural or current bias) and therefore the Lorentz force induces a preferential Josephson vortex flow or intermittent vortex motion. On the other hand, an applied *MW* induces Shapiro steps on the *dc* current-voltage characteristics. From this perspective, understanding how the Lorentz force-driven asymmetric flux motion impacts on either 1) the height of *MW* induced Shapiro resonances or 2) on the array's coherent operation is essential. In the first case it will help minimizing the unwanted interference of Shapiro resonances on the performance of high frequency operation of superconducting devices and their noise spectral density. In the second case, it would be beneficial to our understanding of reaching coherent operation for superconducting *MW* generators/receivers and *FFOs*. Here, we address both these cases as explained in the following. The applied magnetic field *B* penetrates such devices as an ensemble of magnetic flux quanta $\Phi_0$, known as Josephson vortices whose dynamics are very sensitive to applied direct *I* or/and alternating currents $I_{ac}$ (and voltages) originating from *MW* exposure. These devices being inherently asymmetrical to some degree, operate in a non-zero voltage state where *I* produces a Lorentz force that induces a preferential Josephson vortex flow. Thus, the lattice of vortices is moving with a speed (proportional to the measured *dc* voltage *V*) whose magnitude depends on the value of *I*: positive or negative. This phenomenon is called a *B*-field induced magnetic Josephson vortex ratchet effect[12-20]. Josephson ratchets based on very large asymmetric *JJ*-arrays have showed remarkable features such as an ability to amplify the self-induced electromagnetic radiation[19]. Another fundamental phenomenon in *JJ*-based superconducting devices is the so-called inverse *ac* Josephson effect, i.e., the appearance of resonant dc Shapiro steps[21] on the dc current-voltage characteristics (*IVC*) in the presence of an applied *MW* radiation. The Shapiro steps appear at multiple voltages of the ac Josephson effect relation $f=nV/\Phi_0$, with n=1, 2,3,… Surprisingly, the influence of a preferential flux-flow on the strength of the inverse ac Josephson effect, i.e., on the magnitude the Shapiro steps in asymmetrical arrays or on their coherent operation have never been investigated. This is important to fully understand the physics behind the response, coherent operation, and ultimate sensitivity of *JJ*-based devices. Here we report on the influence Josephson vortices flow has on the inverse *ac* Josephson effect in purposely-build asymmetric arrays made of 10 $YBa_2Cu_3O_{7-\delta}$ Josephson junctions (*JJ*) operating in the temperature range (4.2-45) K and irradiated with *MW* radiation in the range (45-75) GHz. We found that for each particular value of *B* a preferential direction of vortex flow dramatically impacts on the appearance of Shapiro steps: if, say, for a positive value of *I* multiple Shapiro steps are well pronounced, then when we reverse the direction of the current, i.e., for -*I*, the Shapiro steps are strongly supressed. Remarkably, when we change the value of the B field slightly the opposite occurs: the Shapiro steps are now strongly supressed for positive *I*, while they are well pronounced for negative *I* (in agreement to general features of dynamical equations which are invariant under B→-B, I→-I, x→-x transforms). Based on extensive numerical simulations we were able to explain this behaviour by various degree of coherence reached in the 10 *JJ*-arrays. Our results are particular relevant where reaching coherent operation in *JJ*-array-based devices is essential such as flux-flow oscillators for sub-terahertz integrated-receivers[19, 22-29] or on-chip superconducting *MW* generators which usually operate for bias currents at the Shapiro step region (suitable for quantum computers[30] and in other applications[29, 31, 32]). Indeed, the Shapiro step is a manifestation of the self-induced locking of the Josephson and resonator dynamics leading to the measurable power emission[21].

The *JJ*-arrays were fabricated by depositing high-quality epitaxial, 100 nm thick *c*-axis oriented $YBa_2Cu_3O_{7-\delta}$ (YBCO) films on 10x10 $mm^2$, 24° symmetric [001] tilt $SrTiO_3$ bicrystals by pulsed laser deposition. A 200nm thick Au layer was deposited in situ on top of the YBCO single-layer film to facilitate fabrication of high-quality electrical contacts for electric transport measurements. Medium-underdoped YBCO films with a critical temperature $T_c$=49 K were subsequently patterned by optical lithography and etched by an Ar ion beam to form asymmetric parallel arrays of 10 *JJs* (see Fig. 1). Within each such parallel array all 10 *JJs* are 3μm wide. The junctions are separated by superconducting loops of identical width of 3μm but variable length. The loops' length increases linearly from 8μm to 16μm in steps of 1μm. Since the individual loop inductances, $L_n$, are proportional to the loop perimeter (1μm corresponds to approximately 1pH) $\beta_{Ln} = 2\pi L_n I_c/\Phi_0$ also increases monotonically by 58% within the 10 *JJs*-array, with n=1, 2, …,9 where $I_c$ is the *JJs* critical current. One can therefore define an average value $\beta=\langle\beta_{Ln}\rangle$ for the array. Namely, $\beta$ can be estimated from both the modulation of $I_c$ with *B* or direct calculations. The bias current *I* is applied symmetrically via the central top and bottom electrodes and *V* is measured across the array. *B* is applied perpendicular to the planar array's structure via a control current $I_{ctrl}$ through an inductively coupled coil. Consequently, an external magnetic flux, $\Phi_{ex}$, is coupled into the array. The *JJ*-array was placed in a Fabry-Perot resonator which was excited at a TEM00k microwaves resonance in the frequency range, *f*, (45-75) GHz at an input power level up to 30 mW[33]. The coupling which was controlled by the rotation of the array relative to the electric field in the waveguide, is minimal when E is parallel to the grid and maximal in the perpendicular direction. Due to the inverse *ac* Josephson effect the applied *MW* induce multiple current Shapiro steps at voltages $V_n$ given by the *ac* Josephson effect relation $V_n = nhf/2e$. Due to the asymmetry of the 10 *JJ*-array for each particular value of *B* the intensity of the Josephson vortex flow along the array (indicated by the horizontal arrow in Fig. 1) for positive *I* is strongly enhanced relative to the case when *I* is negative. This highly preferential Josephson vortex flow in one direction induces a strong asymmetry in the appearance of the Shapiro steps. We fabricated two such devices and both showed a qualitatively similar behaviour. These devices showed a strong and robust *B*-field tunable ratchet effect in the absence of *MW*[20].

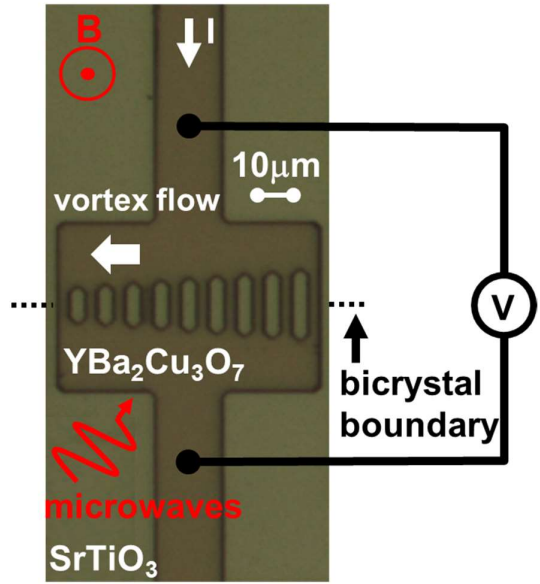

*Fig. 1. Optical micrograph showing an asymmetrical parallel array of 10 YBa$_2$Cu$_3$O$_7$ bicrystal grain boundary Josephson junctions (JJ) fabricated. The Josephson junctions are formed across the bicrystal boundary (shown by dotted line) and can be seen as bridges, grey in colour, crossing it. Within each parallel array the rectangular holes separating the junctions have an identical width of 3 μm and increasing lengths from 8 to 16 μm. A magnetic field B is applied perpendicular to the 10 JJ-array planar structure that is biased with a dc bias current I.*

Families of *IVC's* and *dI/dV's* were measured by a 4 point-contact method at various temperatures *T* between 10K and 49K, for different B field values in the range (-53.5 µT, 53.6 µT) and various *MW* powers (*MW* voltage bias $V_{MW}$ in the range 1-50 mV). The *B* field was changed in small steps of 67 nT. Shapiro steps are better defined on *dI/dV's* relative to the *IV's* data: they appear as peaks as opposed to current steps. For that reason in the following the *dI/dV* data will be analysed. Consequently, we define the height of the Shapiro steps as the height of the Shapiro peak on the *dI/dV* curves. An example of a complete set of data for a particular *MW* bias voltage $V_{MW}$ =4.5 mV is shown in Fig. 2(a): a 3D plot of *dI/dV(V, $I_B$)* extracted from a family of 301 *IVC's* measured at 40 K. Two individual *dI/dV's* for two particular values of the *B* field current $I_B$ are shown in Fig. 2b. $I_c$ and the height of the first Shapiro step versus *B* field current $I_B$ are shown in Fig. 2c. The preferential flux-flow-induced asymmetry in the inverse ac Josephson effect is evident for all values of B field currents measured. Thus, in Fig. 2b, when *B* =9.6 µT the first 5 (n=1-5) Shapiro steps are clearly visible for positive voltages (or current bias *I*) while they are significantly suppressed for negative voltages (or current bias *I*). The asymmetry reverses as we change *B*. Thus, when *B* =17.5 µT, the situation is opposite: the first 5 (n=1-5) Shapiro steps are now well defined for negative voltages (or current bias *I*) while they are significantly suppressed for positive voltages (or current bias *I*). This behaviour qualitatively applies for all *B* field and current values measured: there is always a significant degree of asymmetry in the height of some or all 5 (n=1-5) Shapiro steps for positive *I* relative to negative *I* (see Fig. 2a). For example, the asymmetry of the first Shapiro step between the case of positive *I* relative to the case of negative *I* is shown in Fig. 2c. The lack of correlation between the value of $I_c$ and the height of first Shapiro step is also evident from Fig. 2c. We observed a similar qualitative behaviour to Figs. 2 for many other values of the *MW* bias voltages $V_{MW}$. For a fixed value of *B* the asymmetry stays robust and does not reverse as we change the *MW* power (see Figs. 3a and 3b): that is multiple Shapiro steps are well pronounced for negative *I* only, while no Shapiro steps could be resolved for positive *I*. We observed a similar qualitative behaviour to Figs. 3 for many other values of the *B* field. With increasing *MW* power the amplitude of the Josephson critical current decreases, while the Shapiro steps start developing in increasing order: first the 1$^{st}$ Shapiro step, then the 2$^{nd}$ one, and so on (see Fig.3c). Interestingly, this behaviour is qualitatively similar to that observed for single Josephson junctions [21].

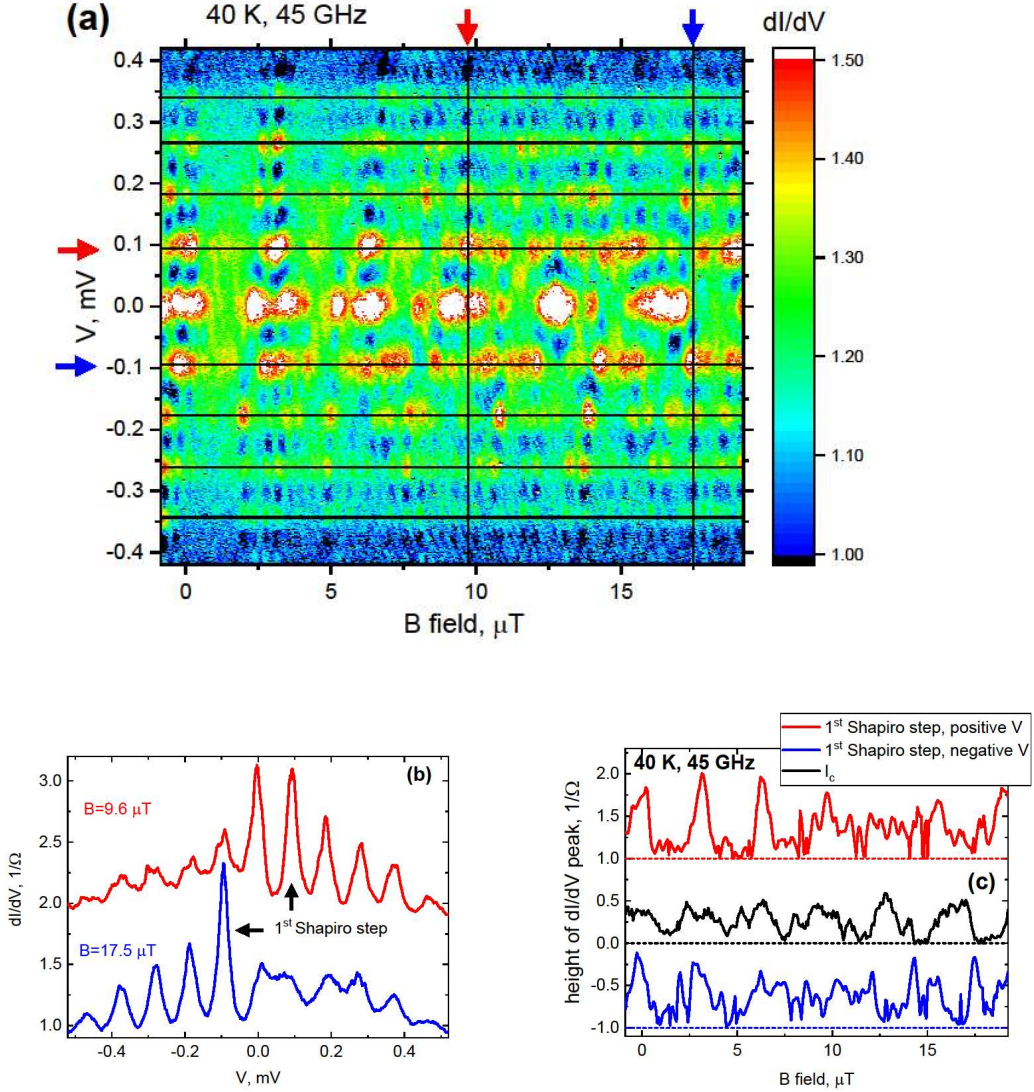

*Fig. 2. Shapiro steps versus B field at a MW bias $V_{MW}$=4.5 mV. (a) False-colour plot of dI/dV(V, $I_B$) extracted from a family of 301 IVC's measured at 40 K for an asymmetric parallel 10 JJ-array for different values of the B. $I_B$ was changed in steps of 67 nT in the range (-0.87, 19.2) µT. The dc bias current I range used in the measurements was (-0.7, 0.7) mA. The white regions centred at V=0 correspond to maxima in the dI/dV due to the array's Josephson junction's critical current. The horizontal grid lines are along the voltage positions of the 8 Shapiro steps voltages $V_n$: ±n×0.094 mV, with n=1, 2, 3, and 4. (b) 2D plots of dI/dV(V) for two different values of the B field: 9.6 µT and 17.5 µT. (c) B field dependencies of $I_c$ and of the height of the first Shapiro step for both negative and negative bias current I extracted from (a). The two 2D plots are cuts of the 3D plots shown in (a) made along lines parallel to the V axis (b) or B axis (c) at the location shown by the red and blue arrows, respectively.*

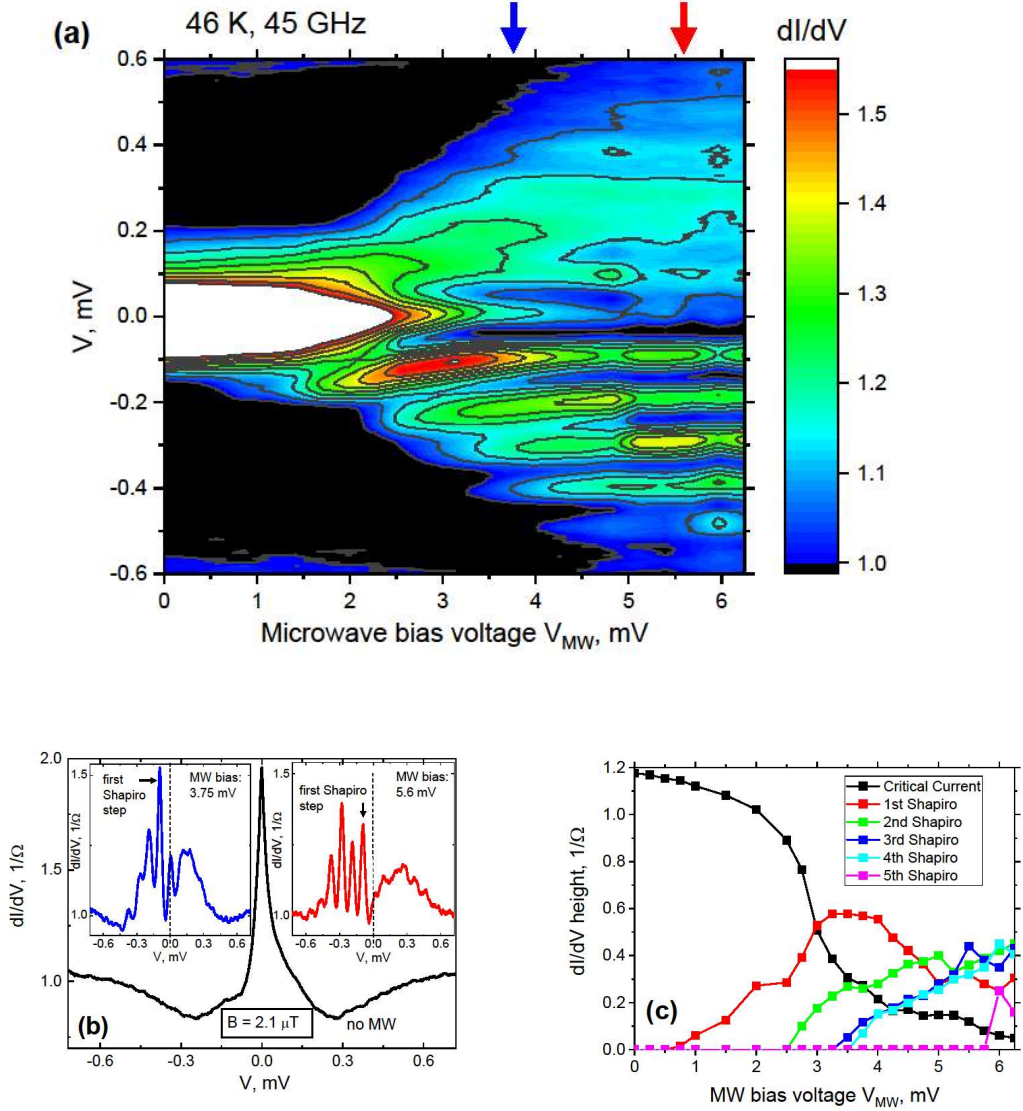

*Fig. 3. Shapiro steps versus MW voltage $V_{MW}$ (power) at a fixed B field of 2.1 µT. (a) 3D plots of dI/dV(V, $V_{MW}$) extracted from a family of 23 IVC's measured at 46 K for an asymmetric parallel 10 JJ-array for different values of $V_{MW}$ in the range (0, 6.2) mV. The dc bias current I range used was (-0.7, 0.7) mA. The large white region centred at V=0 correspond to maxima in the dI/dV due to the array's Josephson junction's critical current. (b) 2D plots of dI/dV(V) for two different values of $V_{MW}$: 3.75 mV and 5.6 mV. These two 2D plots are cuts of the 3D plots shown in (a) made along the lines parallel to the voltage axis at the location shown by the red and blue arrows, respectively. (c) MW power dependencies of the height of dI/dV peaks due to the Josephson critical current and the first 5 Shapiro steps. These 2D plots were extracted from the 3D plots in (a).*

To better understand the physics behind the asymmetry in the Shapiro steps induced by the preferential directions of vortices we perform numerical simulations based on a model consisting of an array of resistively shunted junctions connected via superconducting inductances. The asymmetric Josephson transmission line can be described by a set of coupled ordinary differential equations[20]:

$$\frac{d\varphi_0}{dt} = j - j_{c0}\sin\varphi_0 + \alpha\left[\frac{\varphi_1 - \varphi_0}{A_{\frac{1}{2}}} - h\right] + \sqrt{2D}\,\xi_0$$

$$\frac{d\varphi_n}{dt} = j - j_{cn}\sin\varphi_n + \alpha\left[\frac{\varphi_{n+1} - \varphi_n}{A_{n+\frac{1}{2}}} - \frac{\varphi_n - \varphi_{n-1}}{A_{n-\frac{1}{2}}}\right] + \sqrt{2D}\,\xi_n$$

$$\frac{d\varphi_{n_{max}}}{dt} = j - j_{cn_{max}}\varphi_{n_{max}}\sin\varphi_{n_{max}} + \alpha\left[h - \frac{\varphi_{n_{max}} - \varphi_{n_{max}-1}}{A_{n_{max}-\frac{1}{2}}}\right] + \sqrt{2D}\,\xi_{n_{max}}. \quad (1)$$

In order to match the experimentally implemented design, *n* runs from 1 to 8 in the middle equation of set (1), while $n_{max} = 9$ in the last equation of the set. The dimensionless applied current $j = (I + I_{MW}\cos\omega t)/\langle I_c\rangle$ has a *dc* component *I* and a *MW* component of amplitude $I_{MW}$. Here $\langle I_c\rangle$ is the array's average $I_c$, $\varphi_n$ (with *n=0,…,9*) is the gauge invariant phase difference across the nth junction, dimensionless time *t* is measured in the units of the characteristic relaxation time $\tau = \Phi_0/2\pi cRI_c$ with the flux quantum $\Phi_0$, the speed of light *c*, and the junction resistance *R* (assumed the same for all junctions). The frequency of the applied microwave radiation ω is measured in units of $\omega_c = 2\pi RI_c/\Phi_0$, the Josephson characteristic frequency. The detailed consideration of two-dimensional boundary conditions of the problem suggests the same *j* in all junctions of the line. We also introduce parameter α= τcR/(4πaS$_{1/2}$) with the inter-junction distance *a* and the smallest area $S_{1/2}$ of the array loop between junctions with *n=0* and *n=1*. The array loop area linearly increases with junction number and area ratio $A_{n+1/2}=$ $S_{n+1/2}/S_{1/2} = 1+n/8$, with n+1/2 refers to the array loop between $n^{th}$ and $(n + 1)^{th}$ junctions. The dimensionless magnetic flux $h = HS_{1/2}/\Phi_0$, is measured in units of magnetic flux quantum per smallest array loop area of the array. In the simulations thermal fluctuations have been considered too: we have introduced temporal unbiased δ-correlations Gaussian white noise $\xi_n$ with intensity D, i.e., $\langle\xi_n\rangle = 0, \langle\xi_n(0)\xi_n(t)\rangle = \delta(t)$, which can occur, for instance, due to thermal noise or temporal current fluctuations of the external circuit. We average the time derivative of the gauge invariant phase difference over time and junctions $\bar{\bar{\varphi}}(j) = \langle\langle d\varphi_n/dt\rangle_n\rangle_t$. The dynamical equations (1) used here are similar to the ones previously utilised to describe symmetrical 2D arrays of Josephson junctions[34] investigated experimentally in[35]. However, there is a significant difference in that we consider not only the applied current and the external magnetic field, but also take into account the field gradient trapped in neighbouring holes proportional to the electrical current in the junction connecting these holes. The field gradient and self-induction was ignored in[34], which is key for our analysis of asymmetric Josephson arrays.

A typical set of data is shown in Figs.4a and 4b. Neglecting thermal noise in the simulations (D=0), Shapiro steps are well defined on the IV's and the asymmetric response in the inverse ac Josephson effect is obvious: the first 4 Shapiro steps are clearly more pronounced for positive I, relative to the case of negative I (see main graph in Fig. 4a). When thermal noise is considered (D≠0) Shapiro steps are suppressed and are better visualized on the *dI/dV* curves. Now the similarity with the experiments is striking: compare insets of Fig.3b with the left inset in Fig.4a. The right inset in Fig.4a suggests that the Shapiro steps can be completely supressed for negative *I*, while they are well pronounced for positive *I*. The asymmetry in the Shapiro steps is very significant for all investigated steps n=1,2,3,4 and for the vast majority of values in the entire 2D plane investigated ($i_{ac}$, h). An example is shown in Fig. 4b for the first Shapiro step: n=1. To understand the asymmetrical response in the inverse ac Josephson effect we investigated both the dynamic phase coherence/synchronisation among the 10 junctions in the array, as well as, the static magnetic field distribution $h_i(I)$ along all of the 9 superconducting loops for current biases that correspond to the middle of first Shapiro steps in Fig.4a: *I*=1.05 and *I*=-0.67. It is important to stress that in all simulations performed asymmetric JJ-arrays have always $h_i(I)$ distributions that are highly asymmetric with *I* in high contrast to the case of symmetric *JJ*-arrays. Numerical simulations show (see Fig. 1a in supplementary material) that for a positive bias current of *I*=1.05 there is a high degree of coherence among the array with 7 out of 10 junctions oscillating in-phase leading to a well pronounced first Shapiro step in Fig.

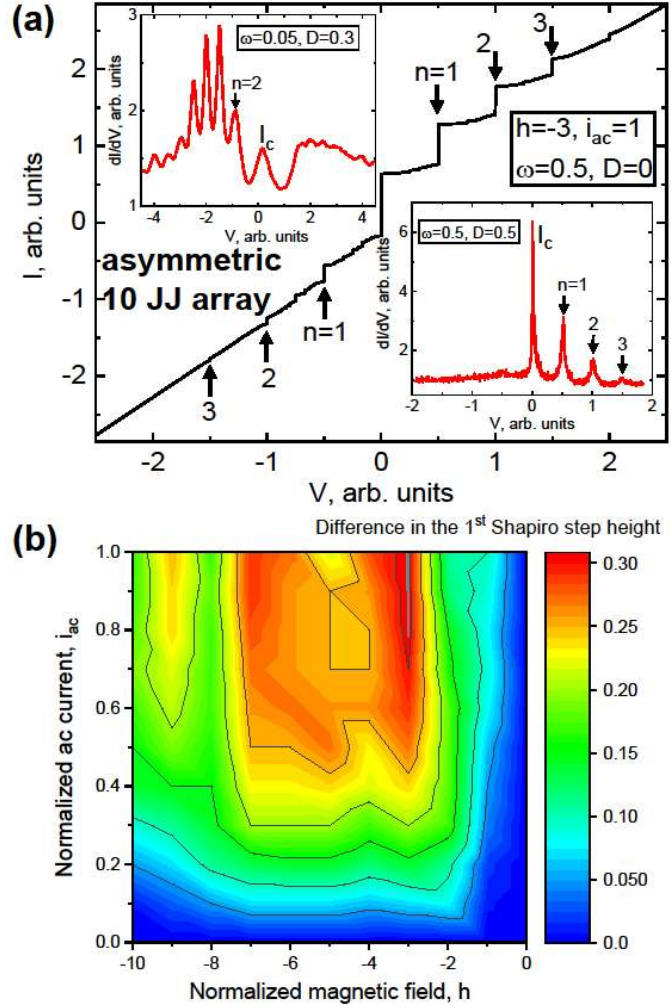

*Fig. 4. Numerical simulations results. a) Electric transport characteristics of the 10 JJ array in the presence of MW of amplitude $i_{ac}=1$ and a magnetic field $h=-3$. Main: IV in the absence of thermal fluctuations (D=0) and frequency $\omega=0.5$. Left inset: dI/dV in the presence of thermal fluctuations (D=0.3) and $\omega=0.05$. Right inset: dI/dV in the presence of thermal fluctuations (D=0.5) and $\omega=0.5$. b) 3D plots of the difference in the first Shapiro steps versus (h, $i_{ac}$) in the absence of thermal fluctuations (D=0).*

4a. In contrast (see Fig. 1b in supplementary material) for a negative bias current of $I=-0.67$ there is a low degree of coherence among the array with no junctions performing in-phase oscillations, leading to a relatively suppressed first Shapiro step in Fig. 4a. Interestingly, there is a strong correlation between the degree of dynamic coherence in the oscillations within the 10 *JJ*-array and the corresponding static flux configuration in the 9 holes at the *I* values corresponding to the Shapiro steps. Thus, for values of $I$ around 1.05 when the first Shapiro step is well pronounced (see Fig.4a), the $B$ field configuration is well structured with significant differences in the values of neighbouring loops for most loops in the array (see Fig. 1c in the supplementary material). Alternatively, we may say that in this case the *JJ*-array is significantly *polarized*, with strong circulating supercurrents around most loops to account for the rather large differences in the $B$ values. This is in high contrast to the case when $I$ take values around -0.67 and the first Shapiro step is strongly suppressed: in this case the $B$ field configuration is *unpolarized*, with small differences in the values of $B$ in most loops and, consequently, small corresponding circulating supercurrents (see Fig. 1c in the supplementary material). It follows that the static flux configuration has a strong influence on the intensity of flux flow previously observed in this particular *JJ*-array [20] which in turns, affects the degree of coherence within the *JJ* array leading to an asymmetric response in the height of Shapiro steps for positive $I$ relative to negative $I$.

We showed experimentally that during the operation of an asymmetrical 10 *JJ*-array placed in an applied $B$ field, the preferential Josephson vortex-flow induces a strong asymmetry in its response to applied *MW* power. Thus, the asymmetry in the inverse ac Josephson effect is evident for all values of $B$ measured: multiple robust Shapiro current steps are formed on the *IVC*'s for a particular positive current bias $I$, but strongly suppressed when we reverse the current direction, i.e., for a bias $-I$. Remarkably, by changing $B$ slightly, the situation is reversed: Shapiro steps are now suppressed for a positive $I$, while well pronounced for a reverse current $-I$ (see Figs. 2). For a fixed $B$ the asymmetry stays robust and does not reverse as we change the *MW* power (see Figs. 3): that is multiple Shapiro steps are well pronounced for negative $I$ only, while no Shapiro steps could be resolved for positive $I$. The extensive numerical simulations performed are in qualitative agreement with the experiments and strongly suggest that the observed asymmetrical response to *MW* is due to a different degree of coherence reached in the *JJ*-array. Our results suggests that vortex motion in asymmetrical devices consisting of either multiple *JJ*s  or a single long *JJ* asymmetrically biased operating under the presence of both a MW and a remnant/applied B field has a fundamental influence in reaching a coherent operation which is essential to consider when designing/fabricating them for various applications such as *FFOs*, flux-driven Josephson (travelling-wave) parametric amplifiers, or on-chip superconducting *MW* generators/detectors.

**Supplementary Material**

See the supplementary material for the details related to JJ-array fabrication and numerical simulations.

Data availability. The data that support the findings of this study are available from the corresponding author upon reasonable request.